# Social Media Monitoring of the Campaigns for the 2013 German Bundestag Elections on Facebook and Twitter

Lars Kaczmirek, Philipp Mayr, Ravi Vatrapu, Arnim Bleier, Manuela Blumenberg, Tobias Gummer, Abid Hussain, Katharina Kinder-Kurlanda, Kaveh Manshaei, Mark Thamm, Katrin Weller, Alexander Wenz, Christof Wolf

Final version 31.03.2014

## Abstract

As more and more people use social media to communicate their view and perception of elections, researchers have increasingly been collecting and analyzing data from social media platforms. Our research focuses on social media communication related to the 2013 election of the German parliament [translation: Bundestagswahl 2013]. We constructed several social media datasets using data from Facebook and Twitter. First, we identified the most relevant candidates (n=2,346) and checked whether they maintained social media accounts. The Facebook data was collected in November 2013 for the period of January 2009 to October 2013. On Facebook we identified 1,408 Facebook walls containing approximately 469,000 posts. Twitter data was collected between June and December 2013 finishing with the constitution of the government. On Twitter we identified 1,009 candidates and 76 other agents, for example, journalists. We estimated the number of relevant tweets to exceed eight million for the period from July 27 to September 27 alone. In this document we summarize past research in the literature, discuss possibilities for research with our data set, explain the data collection procedures, and provide a description of the data and a discussion of issues for archiving and dissemination of social media data.

**Recommended citation**

Kaczmirek, L., Mayr, P., Vatrapu, R., Bleier, A., Blumenberg, M., Gummer, T., Hussain, A., Kinder-Kurlanda, K., Manshaei, K., Thamm, M., Weller, K., Wenz, A., Wolf, C. (2014). Social Media Monitoring of the Campaigns for the 2013 German Bundestag Elections on Facebook and Twitter. Retrieved from http://www.gesis.org/en/publications/gesis-working-papers/

**Acknowledgments**

GESIS initiated the project and was the project lead. The project goals and the conceptualization of the data (corpora) were developed by GESIS, who also undertook data collection on Twitter. The project partner at the Copenhagen Business School, Computational Social Science Laboratory (CSSL), Department of IT Management used their Social Data Analytics Tool (SODATO) to conduct data collection on Facebook.

Lars Kaczmirek was the social science project head and supervisor at GESIS. Philipp Mayr was the computer science project head and supervisor at GESIS. Manuela Blumenberg and Tobias Gummer developed the project goals under the supervision of Lars Kaczmirek. Alexander Wenz helped in researching the politicians and their accounts. Together, they constructed the source information. Armin Bleier and Mark Thamm conducted the data collection on Twitter and developed the necessary tools. Kaveh Manshaei helped with resolving the shortened URL in the Twitter dataset into the original URL. Ravi Vatrapu was the supervisor in Copenhagen. Abid Hussein conducted the data collection on Facebook and developed the necessary tools under the supervision of Ravi Vatrapu. Katharina Kinder-Kurlanda and Katrin Weller developed the perspective on data archiving and dissemination. Katrin Weller contributed the overview of existing research literature. Christof Wolf was the scientific advisor of the project. The project started in March 2013.

# Table of Contents

# 1   Social media research in the political sciences and research goal

Political communication has become a major focus in the growing field of social media studies. Researchers across disciplines and across the globe analyze political online communication with a specific focus on elections – particularly since election campaigns increasingly take place in social media, a process that was prominently recognized during the US election campaign in 2008.

Our goal is to examine various aspects of the communication structures in online media and how such data can add new insights in comparison to existing data from surveys and (traditional) media analyses. By collecting data from both Twitter and Facebook we also add new dimensions to current studies on social media use during elections. Our approach is situated in the broader framework of the German Longitudinal Election Study (GLES), a long term research project that examines the German federal elections in 2009, 2013, and 2017 with the aim to track the German electoral process over an extended period of time (Schmitt-Beck, Rattinger, Roßteutscher & Weßels, 2010). By collecting Twitter and Facebook data about the German Bundestag elections we can supplement traditional research data as used in the GLES. Specifically, the candidate study of the GLES (based on interviews with candidates) can be compared to the actual behavior of candidates on Twitter and Facebook. Similarly, the media corpus analysis of the GLES (analyzing different mass media channels) may be compared with the social media corpus.

Despite a growing research body, there is a lack of shared methods and standards in analyzing electoral races with social media – which means that researchers are still exploring different approaches and that it is not always easy to compare results across studies. In some cases, comparison is difficult due to new and different perspectives on the research topic and due to novel research questions. In other cases, a lack of documented methods leads researchers to developing similar approaches in parallel without profiting from each other's experiences. In addition, data from social media platforms are mostly unavailable for secondary analyses. Traditional publication formats often do not allow researchers to describe processes of data collection and manipulation in sufficient detail. With this paper we try to overcome the lack of documentation of the collection procedures by providing an overview of our applied approaches to collecting data from Twitter and Facebook in order to prepare a dataset for studying the 2013 German Bundestag elections.

Among the platforms that have already been the target of other studies in relation to elections are blogs (Adamic & Glance, 2005; Albrecht, Hartig-Perschke & Lübcke, 2008; Farrell, Lawrence & Sides, 2010), YouTube (Bachl, 2011), Facebook (Williams & Gulati, 2009) and Twitter. Early research on Twitter during elections focused on the US (which is still popular, see e.g., Conway, Kenski & Wang, 2013) and on elections in more instable and also totalitarian systems such as the Iranian elections (Gaffney, 2010) and elections in Ghana (Ifokur, 2010). By now there is also a significant number of case studies for individual countries all over the world, like Sweden (Larsson & Moe, 2012), the Netherlands (Tjong Kim Sang & Bos, 2012), Australia (Bruns & Highfield, 2013; Burgess & Bruns, 2012), India (IRIS & IAMAI, 2013), Canada (Elmer, 2013), Singapur (Sreekumar & Vadrevu, 2013), and South Korea (Hsu & Park, 2012).

Only few studies include comparisons that take into account the interwoven landscape of social media, such as taking into account YouTube links on Facebook (Robertson, Vatrapu & Medina, 2010). Studies that compare results across countries are also rare, although there are first approaches: Larsson and Moe (2014) look at three Scandinavian countries, Nooralahzadeh, Arunachalam and Chiru (2012) compare the US and French elections, and a panel at the Internet Research conference 2013[1] brought together perspectives from four different countries. While these studies look at the context of elections as acute events, others also consider general political discussions (e.g., Highfield, Bruns & Harrington, 2012; Paßmann, Boeschoten & Schäfer, 2014) or the overall setting of e-participation and e-government (Beattie, Macnamara & Sakinofsky, 2012).

The role of online communication and internet technology in German politics has also been studied from different perspectives (Albrecht & Schweizer, 2011; Jungherr & Schoen, 2013; Meckel et al., 2012). The German federal government structure allows to compare elections across the different states, like Elter (2013) has done for seven different German federal state elections. The project "Political Deliberation on the Internet: Forms and Functions of Digital Discourse Based on the Microblogging System Twitter" also monitors several regional as well as the various state elections and analyzes the broader impact of Twitter on political debates in Germany (Thimm, Einspänner & Dang-Anh, 2012). Siri and Seßler (2013) as well as Thamm and Bleier (2013) focus on a set of politicians rather than on selected events like elections. Dang-Xuan et al. (2013) combine the two dimensions and take a closer look at influential individuals during an electoral event in order to investigate emotionality and discussed topics. There are a number of ongoing projects collecting tweets around the 2013 German Bundestag election and some new publications can be expected in the near future.

Previous research has also been inspired by the challenge to use social media to predict election results (e.g., Birmingham & Smeaton, 2011; Tumasjan et al., 2011) which has resulted in a considerable amount of skepticism and some counter examples (e.g., Jungherr, Jürgens & Schoen, 2012; Metaxas, Mustafaraj & Gayo-Avello, 2011). Predictions are a particular case that shows how selected methods and applied modes for data collection (e.g., based upon keywords vs. users, selection of time span) influence the potential outcome of a study. Much more research is conducted, however, not to predict election outcomes but to investigate the roles of politicians, media and publics from various perspectives, for example, by focusing on deliberation and participation. In all cases, however, the chosen methods highly influence what types of conclusions can be drawn. Current approaches comprise quantitative analyses (e.g., number of interactions, network analyses), qualitative analyses (e.g., content analysis of posts) and combined methods – some of them automated, others carried out manually. In all approaches the modes of data collection also have an effect on the scope and limits of the study; if data collection from Twitter is, for example, based on one single hashtag, one needs to be aware that parts of the conversation are lost, as the same topic is likely to be discussed under the heading of different hashtags and not every user includes the same hashtag despite the fact that he or she is referring to the same discussion or topic.

This paper is intended to enable the reader to understand the scope of the available data and to provide an accurate assessment of the potential and the limitations of the data. In the following sections we describe the sources of our data and the rationale of the data collection approach. We then outline the more specific aspects of data generated in Facebook and Twitter together with a short technical description. As much as it is desirable to archive collected data and to make it accessible for secondary analysis to other researchers social media data pose new challenges and questions in this regard. Therefore, we discuss the issues of archiving and dissemination in the last section.

---

[1] https://www.conftool.com/aoir-ir14/index.php?page=browseSessions&form_session=36

## 2　Rationale for building the data set and definition of the corpora

As outlined above the goal of data collection was to collect social media communication which is closely related to the German Bundestag election on September 22nd, 2013. To this end we constructed different data sets which we refer to as the "Facebook corpus of candidates" (corpus 1), the "Twitter corpus of candidates" (corpus 2), the "Twitter corpus of media agents" (corpus 3), the "Twitter hashtag corpus of basic political topics" (corpus 4), the "Twitter hashtag corpus of media topics" (corpus 5), and the "Twitter hashtag corpus about NSA / Snowden" (corpus 6). Corpus 1 includes data collected from the Facebook walls of candidates for the German Bundestag. For the other corpora we collected Twitter data. Corpus 2 is comprised of tweets from candidates for the German Bundestag. Corpus 3 is comprised of tweets from news producers such as journalists. Corpora 4 to 6 contain tweets identified by a list of hashtags which was constructed following a topical approach. Technically, we collected tweets sent from account names of our lists (see below), tweets in which those names were mentioned (i.e., which included the @-prefix) and tweets which matched our hashtag lists (i.e., which included the #-prefix).

In preparation for our collection effort, we identified the most relevant candidates (n=2,346) and checked whether they maintained social media accounts. On Facebook we collected information from 1,408 Facebook walls. The Facebook data was collected in November 2013 for the period of January 1, 2009 to December 31, 2013, thus reaching back to the previous election in 2009 as well. On Twitter we followed a set of 1,009 candidates and 76 other agents, for example, journalists. The Twitter data collection period started on 20 June 2013 and ended on 17 December 2013 (i.e., about 3 months before and after the election), the day Angela Merkel was voted chancellor by the Bundestag and which was the beginning of the new government.

### 2.1　Retrieving the list of Bundestag candidates for Facebook and Twitter

Before actual social media data can be collected, researchers need to decide about the scope of the data corpus. Therefore, we had to construct a list of names of the relevant candidates. This list was the starting point for our search of the social media accounts for both corpus 1 and 2. Relevance was defined as the reasonable likelihood of becoming a member of the Bundestag (see appendix for more details). We refer to this list as the list of candidates although the complete number of overall candidates was higher. The data was collected in a two-stage process.

In the first stage, the names of the Bundestag candidates and details of their candidature (list or direct candidature; constituency) were searched on the webpages of the party state associations (six parties x 16 state associations).[2] If the candidates were not announced online, the names were requested via email or telephone call at the press and campaign offices. Since the direct candidates are elected separately in every constituency and since the party congresses, where the list candidates are elected, take place at different times, our list of candidate names was continuously extended. Although an official list of Bundestag candidates is published by the Bundeswahlleiter (federal returning officer) six weeks before the elections, we decided to investigate the candidate names ourselves. We did this in order to be able to start data collection of social media data simultaneously to the start of the GLES media content analysis in June 2013 and in order to collect data sufficiently in advance before the election would take place.

---

[2] Subsequently, we extended our collection efforts to include the AfD as seventh party (see appendix and section 0).

In the second stage, the Facebook and Twitter accounts of the candidates were identified based on the list of candidates. In addition to the internal Facebook and Twitter search function, the list of social media accounts of current members of parliament on the website pluragraph.de was useful. Furthermore, several of the politicians' or parties' websites linked to their social media accounts.

We applied the following criteria to verify that the accounts were related to the target person: (1) Is a reference to the party, for example a party logo visible? Are Facebook friends and Twitter followers members of this party? (2) Do the candidate's personal or party website link to the profile? (3) Can the candidate be recognized via image or constituency (for direct candidates)? Where available, the verified badge in Twitter was used to select the correct account of a candidate in cases of multiple available accounts.

If the candidate had an account which he or she used for private purposes in addition to his professional account[3], only the professional account was included in our list. During our search for the accounts, this problem occurred primarily with Facebook accounts. Since a list of candidates of the 2009 Bundestag election was already available from the 2009 GLES candidate study, we also searched Facebook accounts for these candidates.

## 2.2 Defining the list of gatekeepers and information authorities for Twitter

Since Twitter is a fast moving medium which takes up and redistributes new information quickly, it is likely that conventional media also use Twitter as a data source. We assume that conventional media select information from Twitter and refine and redistribute the topics over the more conventional media. Corpus 3 was designed to reflect this. We refer to the individuals who would follow such an information gathering approach as "gatekeepers" and searched for them among journalists and editors. In a first step, we identified journalists and editors working in internal political divisions of national daily newspapers and magazines (see appendix) and searched their Twitter accounts. The leading principle in selecting the media sources was whether they were included in the print media content analysis of GLES. The result of this first step is a list of all Twitter gatekeepers of conventional media.

In a second step, we retrieved all accounts that the gatekeepers followed. The assumption behind this approach is that the gatekeepers themselves track what we call "information authorities". The information authorities push topics into Twitter and it is likely that they play a central role in shaping the agenda on Twitter. In order to be counted in the list of information authorities we introduced the criterion that at least 25 percent of the gatekeepers have to follow the account. The list is extended by accounts which are followed by at least 25 percent of the journalists or 25 percent of the editors.

These data may prove useful to supplement research related to both the media content analysis and to all short-term components of the GLES. Furthermore, the communication, bonds and agenda-setting among gatekeepers and information authorities themselves can be the target of research. The gatekeepers and information authorities constitute the source for corpus 3, the Twitter corpus of media agents.

---

[3] We could only identify accounts that were publicly available. We did not search for accounts for which the account holder had decided to make it a "private" account in the sense that it is not shared with the public.

## 2.3 Defining the hashtag lists for Twitter

In defining corpora 4 to 6, the Twitter hashtag corpora, we took an alternative approach which was not restricted to communication around specific Bundestag candidates or journalists. To gain information about the political communication of the population on Twitter, we used thematic hashtags. Here, we defined three procedures which serve to generate three lists of relevant hashtags.

### 2.3.1 Hashtag list 1: Basic political topics and keywords

This list is comprised of the common hashtags (abbreviations) of parties in the Bundestag (see appendix) or of parties which are known to communicate substantially via social media (e.g., the party "Piraten"). The list is complemented with the names of the party top candidates as hashtags (e.g., #merkel). A collection of hashtags for the parliamentary elections in general (e.g., #wahl2013 [#election2013]) completes the list. These hashtags comprise different conjunctions and abbreviations of election, Bundestag, and the year 2013 (see appendix). This list is the source for corpus 4.

### 2.3.2 Hashtag list 2: Media content

This list is based on the coding scheme of the media content analysis of GLES (GLES 2009). Wherever reasonable, one or more hashtags were generated for each code in the coding scheme (e.g., the coding scheme used "Landtagswahl" and the corresponding examples for the hashtags included #landtagswahl, #landtagswahl2013, #landtagswahl13, #ltw). The main challenge in setting up this list was that not all issues could be transformed into meaningful hashtags because topics would become too broad and produce more noise in the data than valuable content. This list is therefore subject to a higher selectivity and less objective than the first list. This list is the source for corpus 5.

### 2.3.3 Hashtag list 3: Case study "NSA / Snowden"

In order to allow a more detailed analysis of the political communication on Twitter for a specific topic we decided to create a third list. This list was constructed to capture the communication around the NSA scandal. Snowden revealed that the NSA has been tapping communication around the world which quickly turned into a wide-spread discussion, generating a very large number of tweets in Germany and abroad. This topic was specific enough and it did not include a wide range of sub-themes and could be covered by a limited number of hashtags. At the same time the issue is discussed extensively by the media and we expect it to be relevant in the future as well. Especially so, since politicians have taken up the issue and further fueled discussion. This third list is also different from the other two in such a way that it was not a predefined list (static approach) but three persons in our research team followed the discussion and added new relevant hashtags as needed. We started with 16 hashtags and decided to add four more between July 22 and 25. This list is the source for corpus 6.

## 2.4 Modifications after the initial setup

With the election the party AfD (Alternative for Germany) made an important leap forward. In the initial concept we had not foreseen these events. Therefore, communication about and from AfD candidates is not initially included in corpus 2 but 15 AfD candidates were added on the 27th of November 2013 to the Twitter data gathering procedure. While it is possible to collect tweets from these accounts back to the start of our data collection efforts, this is not possible for @-messages to these users or tweets including their names as a hashtag. Unfortunately, we are unable to add the Twitter communication for the other corpora because monitoring could only be implemented in real-time making it impossible to capture past events. To keep the data consistent with the overall approach we did not include Tweets from and about AfD in the corpora definition.

Because Facebook posts are more persistent we were able to include data of the candidates of the party AfD. The Facebook walls of AfD candidates for corpus 1 were re-fetched and are part of the corpus definition.

# 3   Facebook data collection

For corpus 1, the Facebook data were collected and analyzed using the purpose-built software application Social Data Analytics Tool (SODATO). This tool allows examining public interactions on the Facebook walls of Bundestag candidates by extracting several conceptual core types of information: Breadth of engagement (on how many Facebook walls do individuals participate); depth of engagement (how frequently do individuals participate); specific analytical issues such as modes of address (measured use of first person, second person, and third person pronouns); the expression of emotion (positive, negative, and neutral sentiment); the use of resources such as webpages and YouTube videos; verbosity; and extent of participation. In the case of modes of address and expression of emotion, one can examine how they evolve over time.

## 3.1   Using SODATO for social data analysis

Social media analytics can be undertaken in at least two main ways – "*Social Graph Analytics*" and "*Social Text Analytics*" (Vatrapu, 2013). Social graph analytics is concerned with the structure of the relationships emerging from social media use. It focuses on identifying the actors involved, the activities they undertake, and the artefacts they create and interact with. Social text analytics is more concerned with the substantive nature of the interactions, it focuses on the topics discussed and how they are discussed: What keywords appear? What pronouns are used? How far are negative or positive sentiments expressed?

These two types of data, we argue, can provide measures of the extent to which the Facebook walls are serving as online public spheres in that:

- The graphical or structural data allow us to map *the breadth* of the public sphere by reporting the overall number of posts made, which of the walls received most posts and whether they linked out to other sources of information. In addition to looking at the posts in the aggregate we can also look at them individually and map cross-linkage across walls. Was the posting entirely independent such that individuals only posted on one wall or did they post more widely on two or three walls?

- The social text data allow us to examine *the depth* of the engagement taking place through the Facebook walls and thus whether walls are acting as an online public space. In particular we look at three key aspects of the posts – their length, their focus in terms of the use of pronouns in the posts – categorizing them as inward (use of 'I') or outward (use of 'you' and 'they'); and the direction of sentiment being positive or negative.

The next section provides a technical description of SODATA (for an earlier version, see Hussain & Vatrapu, 2011). The tool itself can be accessed at http://cssl.cbs.dk/software/sodato/.

## 3.2 Implementing data collection

To fetch the relevant social graph and social text data from the Facebook walls, we used SODATO. SODATO uses and relies on Facebook's open source API named Graph API. SODATO is a combination of web as well as Windows based console applications that run in batches to fetch social data and prepare social data for analysis. The web part of the tool is developed using HTML, JavaScript, Microsoft ASP.NET and C#. Console applications are developed using C#. Microsoft SQL Server is used for data storage and data pre-processing for social graph analytics and social text analytics. A schematic of the technical architecture of SODATO is presented in Figure 1.

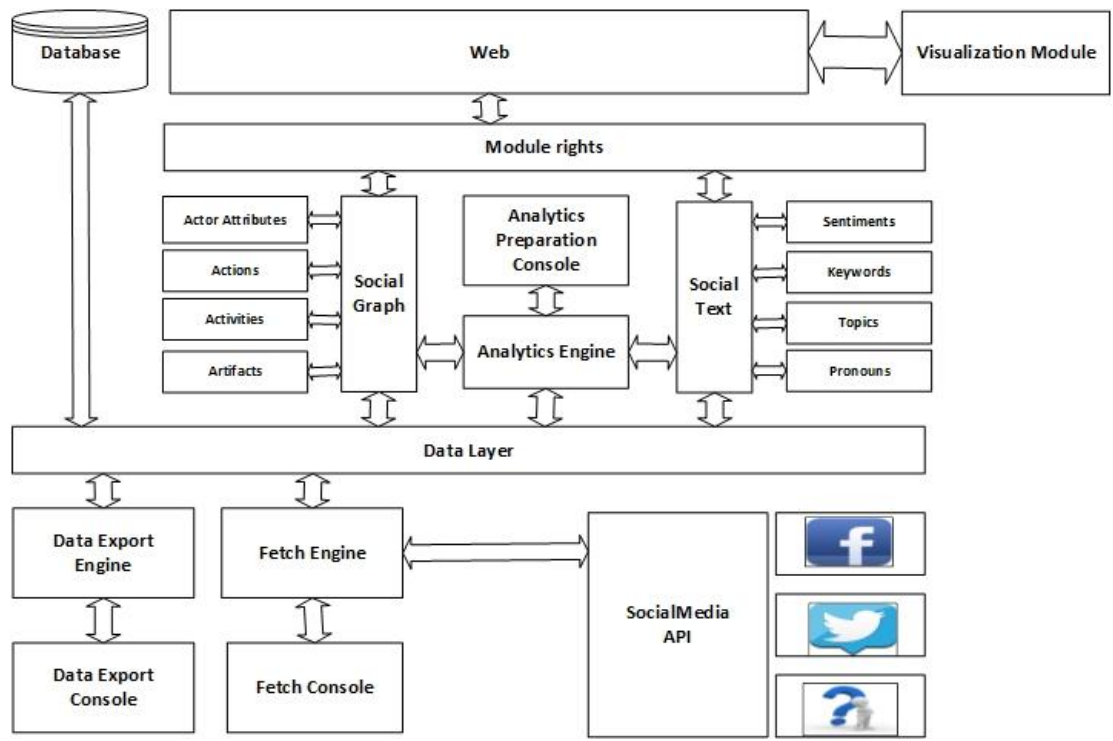

Figure 1: Schematic of the technical architecture of SODATO.

## 3.3   Data structure and Facebook corpus

The data is stored in a MS SQL server in the relational database format. The following attributes are available in the data set:

| Attribute | Description | Example |
| --- | --- | --- |
| *DBITEMID* | unique internal database ID for each action | 00000000 |
| *DBPOSTID* | unique internal database ID for each post (this is the parent with children objects like comments, likes and shares) | 0000000 |
| *FACEBOOKPOSTID* | unique Facebook Graph ID for each post (this is the parent with children objects like comments, likes and shares) | 0000000000_000000000000000 |
| *TIMESTAMP* | Identify the date and time of an action, such as a like | 01/01/2000 00:00 (stored as an internal number) |
| *LASTUPDATED* | Last date and time the post was updated | 01/01/2000 00:00 (stored as an internal number) |
| *EVENTNAME* | type of the Facebook action such as post, comment, share, like | COMMENT |
| *ACTORID* | unique Facebook Graph API ID of the person that performed the action | 00000000 |
| *ACTORNAME* | Facebook username | John Doe |
| *TYPEOFPOST* | type of post such as status, link, video, picture, poll, music, etc. | status |
| *LINK* | the URL if present | www.gesis.org |
| *COMMENTLIKECOUNT* | sum of likes for that particular comment | 4 |
| *TEXTVALUE* | the text | nice comment! |

We collected all available data which was posted on Facebook between January 1, 2009 and October, 31, 2013 to also cover the previous election. Table 1 provides an overview of the size of corpus 1.

Table 1: Size of corpus 1, Facebook corpus of candidates. Showing corpus size for two time periods: (1) January 1, 2009 to October 31, 2013 and (2) 6 weeks before until election day in 2013.

| Feature | 1.1.2009 – 31.10.2013 | 12.08.2013 – 22.09.2013 |
|---|---|---|
| Total accounts identified (walls) | 1,669 | na |
| Successfully fetched | 1,408 | na |
| Failed to fetch | 261 | na |
| Total Posts | 468,914 | 29,782 |
| Total Likes | 3,057,603 | 722,629 |
| Total Comments | 596,569 | 97,911 |
| Total Unique Posters | 18,701 | 2,512 |
| Total Unique Likers | 421,504 | 129,616 |
| Total Unique Commenters | 122,801 | 27,046 |
| Total Unique Actors on all walls | 488,621 | na |

# 4  Twitter data collection

In the following we describe the technical aspects of creating the Twitter corpora. The Twitter monitoring builds up on previous work by Thamm & Bleier (2013). As outlined above Twitter data is used to build five different corpora. Corpus 2 consists of tweets by and addressed to a fixed list of Bundestag candidates. Corpus 3 consists of tweets by gatekeepers and information authorities (all are account names), and corpora 4 to 6 are the result of monitoring different lists of hashtags (see appendix). As data cleaning still needs to be completed, we currently can only preliminary estimate the actual valid size of the corpora. We estimate that the total number of relevant tweets is likely to exceed eight million with over half a million different users.

## 4.1  Implementing data collection

Applying the list of candidate names which have an active professional Twitter account in the 2013 elections we used the Twitter streaming API[4] to receive messages directly from these candidates as well as the retweets of and replies to their messages. We also collected the @messages/mentions and messages which included a hashtag from our lists. For that purpose we developed a software component called TweetObserver that is instantly reading the stream from Twitter resulting from our query in a stable manner. The software needs to register as a Twitter application in order to continuously receive update events for the requested items from the Twitter service. For each account the search query includes the account ID and the name, so that the application is geared towards receiving tweets from a certain account as well as any mentioning of its name. The software was implemented in Java and relied on the Twitter library twitter4j[5]. The software is connected to a MongoDB[6] in which we store the data in JSON format. In the following we describe the data structure of the tweets in the Twitter data set.

---

[4] https://dev.twitter.com/docs/streaming-apis
[5] http://twitter4j.org
[6] http://www.mongodb.org/

The collected tweets are in JSON format and contain at least the following attributes:

| Attribute | Description | Example |
|---|---|---|
| *_id* | tweet ID | 446226137539444736 |
| *userid* | numeric user ID | 630340041 |
| *screenName* | alpha numeric user ID | lkaczmirek |
| *createdAt* | date of tweet | 2014-03-19T11:08:00Z |
| *tweettext* | text of this tweet | @gesis_org is offering #CSES data, providing electoral data from around the world: https://t.co/phtZgGcIjs |
| *hashtags* | internal collection of hashtags with the following attributes | |
| *start* | index of the start-character (the position in the string as a number, the first letter equals index zero) | 23 |
| *end* | index of the end-character (the position in the string as a number) | 28 |
| *text* | the tag itself | cses |
| *mentions* | internal collection of user mentions with the following attributes | |
| *start* | index of the start-character (the position in the string as a number) | 0 |
| *end* | index of the end-character (the position in the string as a number) | 10 |
| *id* | user ID of the mentioned user | 145554242 |
| *screenName* | screen name of the mentioned user (account name) | gesis_org |
| *name* | name of the mentioned user | GESIS |

## 4.2 The Twitter corpora

The data corpus includes tweets sent from account names, tweets with mentions of those names and tweets which matched our hashtag lists. As analysis is still incomplete the statistics below can only describe an arbitrarily defined sub-corpus of our whole data set. We decided to report shortly on a two month period. Table 2 and Figure 2 present examples for unfiltered raw data in the time frame of July 27 to September 27, 2013. Corpus 2 and 3 included 1,176[7] accounts. The number of hashtags used in corpus 4 to 6 amounts to 224.

---

[7] The number of accounts is higher than the pre-specified list of account names due to a list of additional screen names (i.e., handles) which had been kept from a previous, smaller project. Future analyses will exclude this overhead.

Table 2: Example of a portion of unfiltered data between July 27 and September 27, 2013

|  | corpus 2 and 3 (candidates and agents) | corpus 4 to 6 (hashtags) |
|---|---|---|
| number of tweets | 5,573,451 | 3,088,565 |
| number of handles | 356,251 | 181,927 |
| number of unique hashtags | 148,626 | 168,172 |

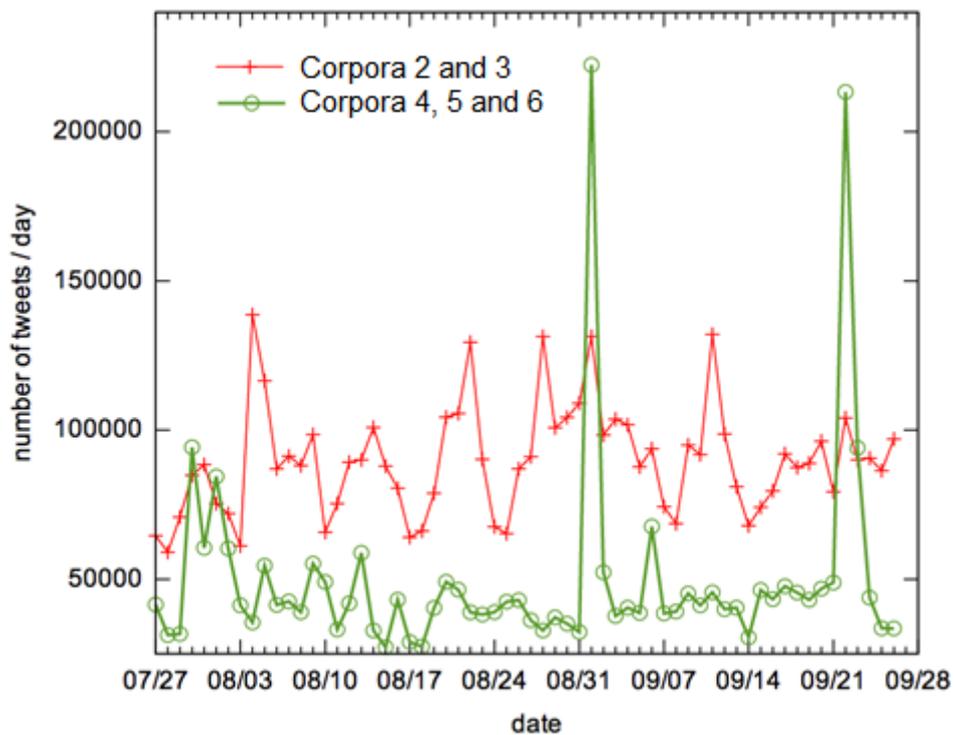

Figure 2: Estimated frequency of the number of tweets per day for the period July 27 to September 27, 2013. The green peaks coincide with the date of the candidates TV debate and the election day.

# 5 Archiving and data distribution

Although various research projects are currently using social media data, and particularly data collected from Twitter, almost no datasets are available for secondary analysis and replication. This is partly due to the complicated legal environment in which social media is situated. The terms of use of companies generating social media data especially add to this effect. In terms of sustainability and verifiability of research it seems desirable, however, to find feasible solutions for long-term archiving and sharing of social media datasets.

Consequently, we outline requirements and conditions for archiving and distribution of social media datasets. For GESIS this means assessing the scope and requirements for archiving a social media dataset and identifying the differences to survey data which the GESIS data archive is used to dealing with and for which we can rely on well-developed tools for archiving, documentation and modes of distribution.

In the following we outline three important areas in which the collected data pose challenges for archiving and distribution.

*Documentation, data structure, tools*: Social media research so far is still in its beginnings and consequently lacks certain standards and methodologies. This applies to both data collection and data analysis. In order to enable re-use of archived datasets, one needs a good description of the available data and the data collection process has to be carefully documented, including the selected tools for gathering the data, data cleaning strategies and data formats. We expect that this document helps to understand what we collected and how we collected our data. We hope that it enables other researchers to compare our approach with others to understand the differences with other datasets that have recently been collected by other projects researching the German election through Twitter or Facebook.

*Data protection and privacy*: Tweets and posts on public Facebook pages are openly accessible to a worldwide audience and users agreed that their posts and tweets are publicly available to a worldwide audience. However, users of social media platforms do not explicitly agree to become a subject of scientific studies. Some users might even expect data transience, especially with tweets. Awareness of these issues is rising among social media researchers (the Association of Internet Researchers AoIR for example has published some guiding advice for internet research)[8]; and ESOMAR has issued a guideline on social media research for market research.[9] Still, comprehensive guidelines for using data from Twitter and Facebook for scientific purposes are not available. User accounts of individuals and institutions of public interest, such as politicians and parties, are often considered to be less problematic than those of 'normal' users, who are less likely to be professionalized in their usage and aware of the publicity they create. Some researchers have decided not to mention user names in publications and to avoid direct quotes from tweets and postings. But there is little discussion on how to handle privacy throughout the process of data collection and storing or on how to apply anonymization strategies effectively. When talking about anonymization, the following two privacy issues are often considered in discussions. The first issue is concerned with tweets/postings that are deleted after the data has been collected in a research project. A user may decide to delete a posting on Facebook or a tweet on Twitter at any time. If the respective tweet has been collected, it would remain in the database if no additional steps are taken to remove it. The second issue concerns user names that are mentioned by other users in tweets/postings. Users may mention other users' names or real names in their posts on social media sites. In the context of political opinions one may find text which could be deemed sensitive in its nature and may require extra consideration. A user might get associated with political discussions, even if he did not actively contribute to them. For example, a @username may be included in a tweet on Twitter like "hey, @username, are you going to vote for party ABC again?". While it might be possible, to filter and anonymize names following the @username structure, it might be more difficult to identify and remove all mentions of names in texts. This makes complete anonymization difficult in social media research.

---

[8] http://aoir.org/documents/ethics-guide/
[9] http://www.esomar.org/uploads/public/knowledge-and-standards/codes-and-guidelines/ESOMAR-Guideline-on-Social-Media-Research.pdf

*Providers' terms of use*: Social media platforms are mostly business corporations that want to protect their interests and maintain certain business models. As such, they might restrict others from using the data in certain ways. Collaborations between archives and social media platforms seem rare. The most notable case is the collaboration of Twitter with the Library of Congress.

Having identified these main challenges for dealing with social media data in the archival processes, it follows that the social media research community would benefit greatly from a workflow for documentation and a concept for data sharing that is in line with social media platforms' terms of service.

Some first ideas on how access might be possible take into account that the GESIS data archive already offers controlled and secure access to sensitive data from surveys via the Secure Data Center (http://www.gesis.org/sdc). Subject to further investigation, such an approach might also be suitable for the use of social media data. A Secure Data Center could be able to resolve requirements of anonymity and data protection. In the Secure Data Center well vetted researchers can work with the data in a protected work environment which is controlled by both technical and organizational safeguards. For example, researchers who want to work with data sign a contract in which they agree not to share the data, not to attempt re-identification of individuals in the dataset and to keep the data secure. Researchers cannot upload or download any data and will perform all their analyses within an encapsulated workspace. All analyses are subject to an output control. This high level of both technical and organizational control allows researchers to work with data that is not fully anonymized. Potential scenarios include enabling access only at a designated computer in a safe room or via a secure remote access – both in combination with signing a special user agreement to follow specific rules.

A different approach that is suggested by Twitter itself has the additional advantage that it deals with some of the challenges of data protection and privacy. Instead of distributing the tweets and its meta-information to a third party researchers are allowed to publish a set of tweet identification numbers (tweet IDs) without any further information (i.e. no tweet contents, no usernames). The main advantage with such an approach is that the tweets themselves are not made available to a third party. Nevertheless, using the Twitter API, tech-savvy users can fetch all necessary information from Twitter and thus reconstruct a dataset on their own. An additional advantage or disadvantage is that the corpus of tweets automatically reflect the privacy settings of the users at the time of the download. Deleted tweets would no longer be included. Yet this method requires technical skills and tools to use the Twitter API and may not be able to include results from certain data cleaning or processing operations. Finally, contractual agreements with social media companies and providers would also solve many of the above mentioned challenges.

# 6 Discussion

Our goal was to monitor the social media sphere as it relates to the German Bundestag election of 2013. To achieve this goal we decided to build six different corpora. For corpus 1, the Facebook corpus of candidates, we monitored how candidates for the German Bundestag present themselves and behave on Facebook. Information on the Facebook walls are fairly persistent in nature; as long as they are not deleted or modified they remain accessible. Therefore, we were able to include data in this corpus which had been generated back during the earlier election in 2009. Nevertheless, Facebook did only receive little attention by politicians in their campaign efforts at that time. The section on Facebook data outlined some of the promising research opportunities with this corpus.

For corpus 2 and 3, the Twitter corpus of candidates and media agents, we monitored how candidates for the Bundestag use twitter and also how gatekeepers (mostly journalists) and information authorities (a prominent set which is followed by the gatekeepers) use Twitter. While historical posts are available on Facebook, the concept of tweeting does not easily allow for long-term access to postings, and especially hinders access to tweets from the past. Technically, we monitored the on-going discussion making it impossible for us to look into the past.

Corpora 4 to 6, the Twitter hashtag corpora, will offer several opportunities for future research. One possibility is to take a single hashtag to try to replicate earlier findings. We see other interesting research possibilities in looking at the subset of hashtags for parties (corpus 4), in comparing the Twitter hashtag corpus of media topics (corpus 5) with the GLES media study or in studying the online discussion with the special set of hashtags for the topic "NSA / Snowden" (corpus 6). The hashtag corpora are probably the most complex ones and many questions need to be addressed concerning data cleaning before we will be able to start working with the data. The main problems arise from the "noise" in the data. As some hashtags are not very specific they refer to other topics unrelated to the political debates we try to monitor and may even be used differently in other countries and languages.

Overall, this paper documents our data collection efforts, the scope and size of the corpora, and some of the problems we encountered in data collection. We also discuss the place of the corpora within the research literature about social media in the political sciences. In an effort to extend and foster such research we conclude with a discussion of the most prominent challenges in data archiving and dissemination. Once these issues are solved, a new phase of social media research can begin as researchers will be able to work on different research questions with validated and agreed upon corpora.

# 8 Appendix

## 8.1 Parties used to compile the list of candidates

The following list contains all parties which were searched for candidates to be included in the corpora. Overall, 2383 candidates were included. Of those, we were able to identify 1669 Facebook accounts which were the source for the Facebook corpus of candidates (also referred to as corpus 1) and 1009 Twitter accounts which were the source for the Twitter corpus of candidates (also referred to as corpus 2).

The list of parties included:

- Christlich Demokratische Union Deutschlands (CDU)
- Christlich-Soziale Union in Bayern e.V. (CSU)
- Sozialdemokratische Partei Deutschlands (SPD)
- Freie Demokratische Partei (FDP)
- BÜNDNIS 90/DIE GRÜNEN (Grüne)
- DIE LINKE (Linke)
- Piratenpartei Deutschland (Piraten)
- Alternative für Deutschland (AfD) [only available in the Facebook corpus]

## 8.2 Sources used to compile the list of gatekeepers and information authorities

The following list contains the names of the newspapers and sources which were searched to compile the list of gatekeepers. We identified 76 gatekeepers and 100 information authorities which were part of the source for the Twitter corpus of candidates (also referred to as corpus 3).

- Bild
- Die Welt
- Die Zeit
- Focus
- Frankfurter Allgemeine Zeitung
- Frankfurter Rundschau
- Spiegel
- Stern
- Süddeutsche Zeitung
- taz

### 8.3 Lists of hashtags

Whenever a tweet included any of the hashtags in our lists as a hashtag (with the prefix #) we aimed to include it in one of the three Twitter hashtag corpora. The hashtags are the source for the Twitter hashtag corpora. The following list contains the 36 hashtags for the basic political topics and keywords (the source for corpus 4):

|  | Topic | Hashtag (#-prefix omitted) |
|---|---|---|
| **Party name** | CDU | cdu |
|  |  | union |
|  | CSU | csu |
|  | SPD | spd |
|  | Bündnis 90/Die Grünen | grüne |
|  |  | grünen |
|  | FDP | fdp |
|  | Die Linke | linke |
|  |  | linken |
|  |  | linkspartei |
|  | Piratenpartei | piratenpartei |
|  |  | piraten |
| **Politics in general** | Politik | politik |
| **Names of top candidates** | Angela Merkel | merkel |
|  | Peer Steinbrück | steinbrück |
|  | Rainer Brüderle | brüderle |
|  | Gregor Gysi | gysi |
|  | Jürgen Trittin | trittin |
|  | Horst Seehofer | seehofer |
|  | Sigmar Gabriel | gabriel |
|  | Philipp Rösler | rösler |
|  | Claudia Roth | roth |
|  | Katja Kipping | kipping |
|  | Bernd Schlömer | schlömer |
| **Election in general** |  | bundestagswahl |
|  |  | bundestagswahl2013 |
|  |  | bundestagswahl13 |
|  |  | btw |
|  |  | btw2013 |
|  |  | btw13 |
|  |  | wahl |
|  |  | wahl2013 |
|  |  | wahl13 |
|  |  | wahljahr |
|  |  | wahljahr2013 |
|  |  | wahljahr13 |

The following list contains the 175 hashtags for the media content keywords (the source for corpus 5):

| Topic | Hashtag (#-prefix omitted) |
| --- | --- |
| Wahlkampf | wahlkampf |
| Wahlprogramm | wahlprogramm |
| Parteiprogramm | parteiprogramm |
| Wahlkampagnen | wahlkampagne |
| Wahlwerbung | wahlwerbung |
| Fernseh-Wahlkampf | wahlwerbespot |
| TV-Duell der Spitzenkandidaten | tvduell |
| TV-Elefantenrunde der Parteivorsitzenden | elefantenrunde |
| Wahlbeteiligung | wahlbeteiligung |
| Direktmandate | direktmandat |
| Überhangmandate | überhangmandat |
| Wahlrecht/Wahlrechtsreform | wahlrecht |
|  | wahlrechts |
|  | wahlrechtsreform |
| Landtagswahl | landtagswahl |
|  | landtagswahl2013 |
|  | landtagswahl13 |
|  | ltw |
|  | ltw2013 |
|  | ltw13 |
| Landtagswahl Bayern | ltwbayern |
| Landtagswahl Hessen | ltwhessen |
| Mitspracherechte der Bürger | bürgerentscheid |
|  | mitsprache |
| Direkte Demokratie/Volksabstimmungen | volksentscheid |
|  | volksabstimmung |
|  | direktedemokratie |
| Politikverdrossenheit | politikverdrossenheit |
| Bundeswehr | bundeswehr |
| Auslandseinsätze | auslandseinsatz |
| Euro-Hawk-Affäre | eurohawk |
|  | drohnen |
| Amigo-Affäre | amigoaffäre |
| Datenschutz | datenschutz |
| Staatliche Überwachung | überwachung |
|  | prism |
| Verkehrspolitik | verkehrspolitik |
| Energiepolitik | energiepolitik |
|  | atomkraft |
|  | atomenergie |

| Topic | Hashtag (#-prefix omitted) |
|---|---|
| | atomausstieg |
| | energiewende |
| | antiatom |
| | antiakw |
| Endlagerung | endlagerung |
| | gorleben |
| Netzpolitik | netzpolitik |
| Umweltpolitik | umweltpolitik |
| | ökosteuer |
| | klimaschutz |
| | klimawandel |
| | hochwasserhilfe |
| Sozialpolitik | sozialpolitik |
| Familienpolitik | familienpolitik |
| | betreuungsgeld |
| | herdprämie |
| | elterngeld |
| Verteilungsgerechtigkeit | einkommensgerechtigkeit |
| | einkommensungleichheit |
| Armut | kinderarmut |
| | altersarmut |
| | einkommensschere |
| Frauenquote | frauenquote |
| Rentenpolitik | rentenpolitik |
| | rentenreform |
| Integrationspolitik | integrationspolitik |
| Gesundheitspolitik | gesundheitspolitik |
| Arbeitsmarktpolitik | arbeitsmarktpolitik |
| | arbeitslosigkeit |
| | jugendarbeitslosigkeit |
| | hartz4 |
| | hartziv |
| | hartz |
| | agenda2010 |
| Lohnpolitik | lohnpolitik |
| | mindestlohn |
| | fachkräftemangel |
| Wirtschaftspolitik | wirtschaftspolitik |
| | wirtschaftslage |
| | finanzkrise |
| | wirtschaftskrise |
| | bankenkrise |

| Topic | Hashtag (#-prefix omitted) |
|---|---|
| Finanzpolitik | ezb |
|  | bankenaufsicht |
|  | bafin |
|  | troika |
|  | finanzpolitik |
|  | staatshaushalt |
|  | staatsdefizit |
|  | steuerpolitik |
|  | steuern |
|  | steuerreform |
|  | finanzmarktsteuer |
|  | finanztransaktionssteuer |
|  | steuergeschenke |
|  | solidaritätszuschlag |
|  | soli |
|  | länderfinanzausgleich |
| Bildungspolitik | bildungspolitik |
|  | schulpolitik |
|  | hochschulpolitik |
|  | forschungspolitik |
|  | kulturpolitik |
| Politiker allgemein | politiker |
| Bundespräsident | bundespräsident |
|  | gauck |
| Bundesregierung | bundesregierung |
|  | bundesreg |
| Bundesminister | bundesminister |
|  | minister |
|  | westerwelle |
|  | friedrich |
|  | leutheusser |
|  | schäuble |
|  | rösler |
|  | vonderleyen |
|  | aigner |
|  | maiziere |
|  | bahr |
|  | ramsauer |
|  | altmaier |
|  | johannawanka |
|  | niebel |
|  | pofalla |

| Topic | Hashtag (#-prefix omitted) |
|---|---|
|  | schroeder_k |
| Bundestag | bundestag |
| CDU/CSU-Fraktion | cdufraktion |
|  | csufraktion |
| SPDFraktion | spdfraktion |
| Bündnis 90/Die grünen-Fraktion | grünefraktion |
| FDP-Fraktion | fdpfraktion |
| Piraten-Fraktion | piratenfraktion |
| Koalitionen allgemein | koalition |
| Große Koalition (prospektiv) | großekoalition |
| Rot-Grüne Koalition (prospektiv) | rotgrün |
| Rot-Gelb-Grüne Koalition (prospektiv) | ampelkoalition |
| Schwarz-Gelbe Koalition (aktuell) | schwarzgelb |
| Schwarz-Grüne Koalition (prospektiv) | schwarzgrün |
| Schwarz-Gelb-Grüne Koalition (prospektiv) | jamaikakoalition |
|  | schwampel |
| Bundesrat | bundesrat |
| Landtag | landtag |
| Bundesverfassungsgericht | bundesverfassungsgericht |
|  | verfassungsgericht |
|  | bverfg |
|  | verfassungsrichter |
| Ministerpräsidenten | kretschmann |
|  | seehofer |
|  | wowereit |
|  | platzeck |
|  | böhrnsen |
|  | olafscholz |
|  | bouffier |
|  | sellering |
|  | stephanweil |
|  | hannelorekraft |
|  | dreyer |
|  | krampkarrenbauer |
|  | tillich |
|  | haseloff |
|  | albig |
|  | lieberknecht |
| Junge Union (JU) | jungeunion |
| Jungsozialisten in der SPD (Jusos) | jusos |
| Grüne Jugend | grünejugend |
| Junge Liberale (JULis) | jungeliberale |

| Topic | Hashtag (#-prefix omitted) |
|---|---|
| Linksjugend (solid) | linksjugend |
| Parteitag | parteitag |
| Infratest dimap | infratestdimap |
| | infratest |
| TNS emnid | tnsemnid |
| FORSA | forsa |
| Forschungsgruppe Wahlen | forschungsgruppewahlen |
| | politbarometer |
| Institut für Demoskopie Allensbach | allensbach |

The following list contains the 20 hashtags for the case study "NSA / Snowden" (the source for corpus 6):

| Topic | Hashtag (#-prefix omitted) |
|---|---|
| Edward Snowden | snowden |
| | snowdenasyl |
| | snowstorm22 |
| | Snowden_to_Germany |
| NSA | nsa |
| Prism | prism |
| | antiprism |
| Tempora | tempora |
| XKeyscore | XKeyscore |
| Discussion in Germany | abhörskandal |
| | spionage |
| | bda |
| | vds |
| | gegenvds |
| | uanm |
| | bnd |
| | supergrundrecht |
| | stopwatchingus |
| Other | Whistleblower |
| | YesWeScan |